\documentclass[10pt,conference]{IEEEtran}

\IEEEoverridecommandlockouts

\usepackage{cite}
\usepackage{amsmath,amssymb,amsfonts}
\usepackage{algorithmic}
\usepackage{graphicx}
\usepackage{textcomp}
\usepackage{xcolor}
\usepackage{url}
\usepackage{booktabs}
\usepackage{fontawesome}
\usepackage{flushend}
\usepackage{fp}

\definecolor{jccolor}{rgb}{0.1,0.7,0.8}
\definecolor{vlcolor}{rgb}{0.9,0.1,0.1}
\definecolor{cpcolor}{rgb}{0.3,0.3,0.7}
\definecolor{ascolor}{RGB}{163,96,50}
\definecolor{ahcolor}{rgb}{0.36, 0.54, 0.66}

\newcommand{\surveynumber}[1]{(\faBarChart\,#1\% adoption)}
\newcommand{\surveynumberchallenge}[1]{\faBarChart\,#1\%}
\newcommand{\allsurveynumbers}[6]{%
 \FPeval{\sumadopted}{clip(#1 + #2 + #6)}%
 \FPeval{\sumliked}{clip(#1 +#2)}%
 \faBarChart\ \sumadopted \% adopted, \sumliked \% satisfied, #6\% did not find useful}
\newcommand{\participantlist}[1]{\faComments\,#1}
\newcommand{\interviewquote}[2]{\faQuoteLeft\,#1:\textit{``#2''}}

\newcommand{\challenge}[1]{\faTable\,#1}

\begin{document}

\title{Beyond the Comfort Zone: Emerging Solutions to Overcome Challenges in Integrating LLMs into Software Products}

\author{\IEEEauthorblockN{Nadia Nahar,\IEEEauthorrefmark{1}\IEEEauthorrefmark{2} 
Christian Kästner,\IEEEauthorrefmark{2} Jenna Butler,\IEEEauthorrefmark{3} Chris Parnin,\IEEEauthorrefmark{3} Thomas Zimmermann,\IEEEauthorrefmark{3} Christian Bird\IEEEauthorrefmark{3}}
\IEEEauthorblockA{\IEEEauthorrefmark{2}Carnegie Mellon University, \IEEEauthorrefmark{3}Microsoft Research\\
\IEEEauthorrefmark{1}nadian@andrew.cmu.edu
}
}

\maketitle

\begin{abstract}
Large Language Models (LLMs) are increasingly embedded into software products across diverse industries, enhancing user experiences, but at the same time introducing numerous challenges for developers. Unique characteristics of LLMs force developers, who are accustomed to traditional software development and evaluation, out of their comfort zones as the LLM components shatter standard assumptions about software systems. This study explores the emerging solutions that software developers are adopting to navigate the encountered challenges. Leveraging a mixed-method research, including 26 interviews and a survey with 332 responses, the study identifies 19 emerging solutions regarding quality assurance that practitioners across several product teams at Microsoft are exploring. The findings provide valuable insights that can guide the development and evaluation of LLM-based products more broadly in the face of these challenges.
\end{abstract}


\footnotesize{
\vspace{10pt}
\noindent\textit{``It's a big unknown that makes me very uncomfortable. It keeps me up.''}
\vspace{-5pt}
\begin{flushright}
    -- Engineer on integrating LLMs into their product
\end{flushright}
}

\normalsize

\section{Introduction}

Large language models (LLMs) have received massive attention and have impacted various industries across all kinds of applications. Beyond generic chatbots such as ChatGpt~\cite{openai} or Bing Copilot~\cite{bing}, LLMs are now integrated as features in a wide array of products and services in domains as varied as healthcare~\cite{health1,health2,health3}, legal~\cite{legal1,legal2}, and sales~\cite{sales}. Leading tech giants such as Microsoft, Google, Amazon, and Apple have all have all embraced LLM technology and have made AI a central part of the user experience of their software products~\cite{report}. For instance, Microsoft incorporates LLMs into its Office suite to offer advanced functionalities such as automatically generating slides in PowerPoint or assisting users in composing email responses in Outlook. We refer to products that integrate LLMs to provide application features as \textit{LLM-based products} in this paper.

Incorporating LLMs into software products introduces disruptions and challenges to established workflows and traditional software-engineering practices -- \textit{pushing developers used to traditional software products out of their comfort zones.} Among others, LLMs introduce new failure modes, shift manual efforts to different tasks such as creating and labeling test data, and shift or introduce additional concerns for latency, cost, energy consumption, fairness, reliability, and compliance. Prompt engineering emerges as a new skill and building complex prompt pipelines introduces another layer of complexity~\cite{liang2024prompts,parnin2023building}. Practitioners struggle particularly with adjusting to new forms of quality assurance for LLM-based features, given a lack of clearly established testing processes and a significant degree of subjectivity -- for example one of our interviewees remarked \textit{``The hardest thing has been [answering] `What is a bug?' Like we have gotten into so many arguments [...].''}

While researchers have made significant efforts to comprehend the \textit{challenges} associated with building machine-learning-based products generally (see a recent survey \cite{nahar2023meta}) and LLM-based products specifically~\cite{hassan2024rethinking,dolata2024development,parnin2023building}, efforts to identify,  catalog, and evaluate \textit{emerging solutions} -- whether in the form of \textit{tools}, \textit{techniques}, and \textit{(best) practices} -- have been fragmented. 
There are many lists collecting various  \textit{LLMOps} tools, with many startups competing in this field to aiding prompt engineering, prompt optimization, monitoring, evaluation, and other development, maintenance, and operations tasks~\cite{LLMOps1,LLMOps2}. There are also academic papers proposing diverse methods and techniques as solutions to support various aspects of development and evaluation of LLM-based products, e.g.,~\cite{kim2024evallm,mishra2023promptaid,strobelt2022interactive,reif2023visualizing,shankar2024spade,rebedea2023nemo}. At the same time, there are increasing concerns about some common practices, such as using LLMs to validate the output of LLMs \cite{shankar2024validates,Murugadoss:EvaluatingTheEvaluator:2024}. Despite the proliferation of these tools and methodologies and lots of (often controversial) opinions about them, their adoption and effectiveness in real-world settings is not well-documented. 

With a mixed-method study combining 26 interviews and a survey with 332 responses~\cite{o2010assessing,creswell2003advanced}, we explore what \textit{emerging solutions} practitioners across many product teams at Microsoft have adopted for common LLM-related challenges and what solutions they are exploring. Our study largely confirms the known challenges (cf. Table~\ref{Table:challengesEval}) and, more importantly, uncovers that several emerging solutions, including \textit{combining qualitative and quantitative metrics to evaluate LLM outputs}, \textit{using LLM-as-a-judge to evaluate subjective metrics}, and \textit{establishing extensive guardrails} are already broadly adopted, whereas others, such as \textit{setting up end-to-end test automation} are adopted by some teams.  In a nutshell, we attempt to answer the following research question: \textbf{What {emerging solutions} (tools, techniques, and practices) are the practitioners adopting to overcome LLM-related challenges?} We identified 19 emerging solutions related to quality assurance, which is the focus of this paper, and seven more regarding development and prompting, reported in the appendix\footnote{The appendix will be attached to the paper once it is published.}. Given the broad portfolio of products and large community of software practitioners at Microsoft, this paper offers an unique perspective on a variety of experiences and approaches encountered while building LLM-based products. We believe these insights can contribute real-world insights that can inform the development and evaluation of LLM-based products more broadly.

In this paper, we make the following contributions:
\begin{itemize}
    \item \textbf{Confirmation and Expansion of Disruptions:} We provide a comprehensive investigation of the disruptions practitioners face when integrating LLMs into software products, confirming existing literature on LLM-related challenges while uncovering new, underexplored disruptions that developers encounter.
    \item \textbf{Identification of Emerging Solutions:} We present 19 emerging solutions that developers are adopting to address these disruptions, offering concrete insights into how these solutions are enabling engineering teams to accommodate LLM-specific issues.
    \item \textbf{Validation Through a Mixed-Methods Approach:}  Beyond the exploratory interviews, we conducted a large-scale survey to quantify the prevalence and effectiveness of the identified disruptions and solutions. This mixed-methods approach strengthens our findings by combining qualitative insights from interviews with quantitative validation from survey data, increasing the generalizability and impact of our results.
\end{itemize}

\section{Challenges and Solutions for Building LLM-Enabled Products (Related Work)}

\subsection{LLM-based Products} 
In recent years, machine learning generally and LLMs specifically have attracted widespread attention, enabling developers to build products around these models~\cite{dolata2024development,parnin2023building,christianbook}. With pre-trained LLMs from vendors like OpenAI or Meta, typically accessed via APIs but possibly also deploying ``open'' models locally, developers can customize LLMs for specific tasks by developing dedicated prompts. The integration of LLMs can be simple as prompt engineering lends itself to rapid prototyping~\cite{jiang2022promptmaker,wu2022promptchainer,liu2022design}, but can also use sophisticated, multi-step pipelines and integrations with information retrieval systems~\cite{lewis2020retrieval,topsakal2023creating,jeong2023generative}. In this paper, we refer to products incorporating these features as \textit{LLM-based products}.

\subsection{Known Challenges in Building LLM-Enabled Products}

After researchers have extensively studied challenges in developing machine-learning solutions and integrating them into software products, see a recent survey~\cite{nahar2023meta}, more recently, researchers have also explored challenges specifically for integrating LLMs into software products by engaging with industry practitioners and freelancers, e.g.~\cite{dolata2024development,hassan2024rethinking,parnin2023building,liang2024prompts}. Focusing on quality assurance problems, we summarize the key challenges identified in these studies in Table~\ref{Table:challengesEval}. As part of our own interviews and survey in this research, we found largely the same challenges, which both replicates and confirms prior work and additionally provides confidence that our study of solutions is conducted in an environment that faces the same challenges as the larger community.

\subsection{Proposed and Emerging Solutions} 

There is a vast number of \textit{proposed solutions} -- tools, techniques, and practices -- suggested by researchers and practitioners. These might be suggested in academic papers (e.g., user interfaces to surface ethical concerns during prompt development~\cite{wang2024farsight}), popular blog posts (e.g., suggesting testing strategies for LLMs grounded in property based testing~\cite{Ribeiro2023-rw}), and startups promoting their services (e.g., tooling to validate LLM outputs~\cite{predictionguard}). Many of these solutions are presented under the label \textit{LLMOps}.

Given the vast number of solutions suggested through academic papers across many venues, all kinds of open-source projects and proprietary tools and services, many of which may have received little evaluation or adoption, we do not attempt a comprehensive survey of \textit{proposed solutions} but refer the interested reader to lists of LLMOps tools~\cite{LLMOps1,LLMOps2} and recent surveys of research on prompt engineering~\cite{sahoo2024systematic,chen2023unleashing,vatsal2024survey} and various examples of research automatically or interactively optimizing prompts~\cite{dspy,shin2020autoprompt,pryzant2023automatic} and evaluating prompts~\cite{kim2024evallm,dixit2024retain,strobelt2022interactive,jiang2022promptmaker,wu2023scattershot,petridis2024constitutionmaker}. Instead of a comprehensive survey, our research focuses on identifying \textit{emerging solutions} that are increasingly adopted and sometimes shared as best practices within Microsoft.

\newcommand\rowspacing{6pt}
\begin{table*}[!htbp]
\caption{Challenges of evaluating LLM-based products}
\label{Table:challengesEval}
\begin{tabular}{@{}p{8cm}p{9.5cm}}
\toprule
\textbf{Known and Confirmed Challenges} & \textbf{Our Participants' Views on the Challenges}
\\ \midrule

\challenge{C1}: \textbf{Lack of specification}.
The definition of a bug is ambiguous and debatable~\cite{christianbook,gero2022sensemaking}. & \surveynumberchallenge{29.6} \participantlist{P9, P10, P14} 

\interviewquote{P14}{How do you test these things are doing well or not doing well? What's the bar.}\\[\rowspacing] 

\challenge{C2}: \textbf{Subjectivity}.
Determining expectations and what fits as correct answers is challenging~\cite{shankar2024validates,aroyo2015truth,mishra2023promptaid}. & 
\surveynumberchallenge{36.7} \participantlist{P2, P9, P10, P14, P16, P17, P24}

\interviewquote{P17}{So in these cases it's more subjective to even measure the quality on the output, and figuring out the rubrics.} \\[\rowspacing] 

\challenge{C3}: \textbf{Metrics dilemma}.
Developing the right metrics set is complicated~\cite{shankar2024spade,dolata2024development}. 
How do we know which metrics to use, how to measure them, and if the chosen metrics are appropriate? 
\begin{itemize}
    \item[(a)]The selection of metrics tends to be \textit{ad hoc}. 
    \item[(b)] Many teams rely on common known metrics without considering whether their \textit{specific use case fits} to them. 
    \item[(c)] Defining \textit{suitable measurements} for metrics is difficult because it’s subjective and not straightforward. 
    \item[(d)] There is a significant \textit{disparity between offline and online metrics}. Comparable sets of metrics for both offline and online evaluation are necessary, but currently lacking.
    \item[(e)] Online metrics provide \textit{weak signals} and offer little insight into output quality.
\end{itemize} &

\interviewquote{P23}{That's really hard to map back to some of these really core metrics [...] So I think that's sometimes like more art than science.}

(a): \surveynumberchallenge{46.5} \participantlist{P5, P17, P23, P25}

\interviewquote{P17}{People just kind of end up doing what they need to do in an ad hoc way.}

(b): \surveynumberchallenge{31.2} \participantlist{P5, P7, P20}

\interviewquote{P5}{We just pick the ones that we are interested in and we just run the test.}

(c): \surveynumberchallenge{46.5} \participantlist{P1, P4, P7, P10, P15, P16, P26}

\interviewquote{P1}{Measurement of the offer is another big technical challenge.}

(d): \surveynumberchallenge{49.7} \participantlist{P1, P26}

\interviewquote{P26}{Our pain points right now are more around metrics and like representativeness of offline data [...] you get these offline metrics that they will translate to something and online like that's a big gap.}

(e): \surveynumberchallenge{30.3} \participantlist{P8, P22, P26}

\interviewquote{P8}{I don't think we have any kind of signal which will trade that we are doing an awesome job on generating content.}\\[\rowspacing] 



\challenge{C4}: \textbf{LLM properties}.
LLMs are made non-deterministic; teams struggle with testing, as outputs are not consistently reproducible. LLMs often hallucinates, making them unreliable~\cite{shankar2024validates,dolata2024development,parnin2023building,LLMsurvey}. &

\surveynumberchallenge{57.7} \participantlist{P1, P2, P3, P4, P5, P7, P9, P10, P12, P13, P14, P15, P17, P18, P22} 

\interviewquote{P9}{How do we deal with this nondeterminism problem?} 

\interviewquote{P10}{It's gonna hallucinate sometimes. It's gonna not do what you ask it to do.}\\[\rowspacing]

\challenge{C5}: \textbf{Lack of robust evaluation methods and pipeline}~\cite{zamfirescu2023johnny,dolata2024development,parnin2023building}. 

\begin{itemize}
    \item[(a)] Many teams \textit{lack proper evaluation mechanisms} or have \textit{inconsistent evaluation practices}. Unit testing prompts are difficult and there are no effective methods for evaluating non-deterministic models beyond basic health checks.
    \item[(b)] Teams report the absence of a cohesive \textit{evaluation pipeline}, having separate evaluations on different platforms with no integration.
    \item[(c)] Evaluations using LLMs are unreliable, frequently requiring multiple attempts to determine whether an issue is due to \textit{flakiness} or another factor. 
    \item[(d)] The sheer number of \textit{variables} involved in testing complex prompts makes it hard to analyze all dimensions and user inputs confidently. 
    \item[(e)] Creating test cases depends heavily on \textit{synthetic data}, often generated using LLMs themselves, as access to real data is limited.
    \item[(f)] Numerous instances of bugs reaching production highlight \textit{inadequate testing} and overlooked \textit{edge cases}.
\end{itemize}
 &  
 
(a): \surveynumberchallenge{36.3} \participantlist{P2, P5, P6, P11, P12, P16, P20}

\interviewquote{P12}{We are working on building more complete test suite because current test suite is still very limited.}

(b): \participantlist{P1, P2, P6, P15}

\interviewquote{P2}{We don't have a strong evaluation pipeline right now.}

(c): \surveynumberchallenge{65.6} \participantlist{P6, P7, P9, P16}

\interviewquote{P6}{It's been really flaky to where it'll fail a lot of the time.}

(d): \surveynumberchallenge{52.0} \participantlist{P1, P2, P4, P6, P7, P9}

\interviewquote{P4}{It's like one prompt. We want to be able to do that as one evaluation, and I'm still looking for where I can run that automatically so that I could get both the breakdown of scores and the single score. There can be pros and cons to it, like maybe we want to run it all at once, or separate it but also giving the context. That's gonna give all these dimensions [that] can help it score and give more. Like more accuracy for each specific score when it knows the other dimensions. Uh. And so we're still like kind of playing around with which one is the right type of evaluation. And there's lots more.}

(e): \surveynumberchallenge{42.1} \participantlist{P12, P15, P16, P25}

\interviewquote{P15}{A big challenge is like getting right data to test these LLMs. Like across everything. So creating synthetic data, [trying] understanding like what they can [look] like.}

(f): \participantlist{P2, P3, P4, P5, P6, P11, P14, P15, P16, P19}

\interviewquote{P5}{Edge cases where [LLM feauture] just actually blurbs out it's entire command} \\[\rowspacing]

\challenge{C6}: \textbf{Manual efforts}.
Manual testing is common but labor-intensive and inefficient. Some teams avoid automation to move quickly but later experience drawbacks.~\cite{mishra2023promptaid,liang2024prompts} &

\surveynumberchallenge{76.6} \participantlist{P1,P2, P9, P10, P11, P12, P14, P15}

\interviewquote{P9}{this process is extremely manual, right? It is not possible to fully automate it because it requires us to keep our brain on.}\\[\rowspacing]

\challenge{C7}: \textbf{Infrastructure constraints}.
There is an over-dependence on the current evaluation infrastructure that lacks flexibility for all use cases. The causes of test failures are not clear, documentation is sparse, and communication is a burden. In some instances, responsiveness and support are inadequate~\cite{LLMsurvey}. & 
\surveynumberchallenge{43.1} \participantlist{P1, P2, P4, P6, P7, P16, P25, P26} 

\interviewquote{P6}{Kind of been challenging because [infrastructure] is owned by another team [...] Nobody on our team really has access to that or fully understands how it's powered and we run into problems with it all the time. So like the data, we'll be looking at the dashboard and like we don't have data for the past two weeks [...]. And then we have to go find someone and poke them and be like, hey, what's going on?}\\[\rowspacing]

\challenge{C8}: \textbf{Model migration issues}.
Evaluating models’ post-migration is problematic; minor tweaks can cause substantial changes, complicating regression testing~\cite{liang2024prompts,ma2024my,sclar2023quantifying}. & 
\surveynumberchallenge{40.4} \participantlist{P12, P14, P16}

\interviewquote{P12}{If you like migrate the model or any changes in the prompts, you have to go through everything again.}\\[\rowspacing]

\challenge{C9}: \textbf{Compliance}.
Compliance processes are lengthy, manual, and bureaucratic, involving excessive paperwork and manual effort~\cite{parnin2023building}. & 

\surveynumberchallenge{30.7} \participantlist{P1, P2, P3, P5, P7, P8, P11, P12, P13, P14, P15, P16, P23}

\interviewquote{P13}{That's (compliance) like a huge, huge challenge and I think just in general, you know, Microsoft is super careful with customer data, which obviously is good.}\\

\bottomrule
\end{tabular}
\begin{center}
\vspace{-4pt}
\scriptsize{ \faBarChart: \% of responses from survey indicating that the challenge is `hard' or `extremely hard;' \faComments:~Interview participants reporting the challenge; \faQuoteLeft: Representative interview quotes}
\end{center}
\end{table*}

\section{Research Design}
We employed the established sequential exploratory mixed-methods research design~\cite{creswell2003advanced} that performs research in two phases: In a first qualitative phase we explore challenges and emerging solutions through interviews. Then, in a subsequent second phase, we quantify the observed challenges and emerging solutions through a survey with a much larger sample size. This enables an in-depth open-ended exploration of emerging themes beyond a predefined list but also allows us to provide quantitative insights about the prevalence of these emerging solutions in the studied teams at Microsoft.

\subsection{Interviews}
In the first research phase, we created an interview guide to broadly explore the \textit{current state}, \textit{challenges}, and \textit{emerging solutions} across the entire lifecycle of LLM-enabled products and features. After feedback from experts and a pilot test, we refined the guide, emphasizing the \textit{evaluation} phase and adding relevant questions to thoroughly explore this topic.

To recruit a varied and representative group of interviewees aligned with our research goals, we used \textit{purposive sampling}~\cite{campbell2020purposive}, explicitly recruiting interviewees from different teams from different departments within Microsoft that had integrated LLM components into their products for various use cases. In addition, we aimed to  recruit multiple team members in different roles (managers, engineers, and data scientists) from each product team to explore different perspectives. In total, we conducted four pilot interviews, followed by 26 interviews we analyzed for this paper (cf. Table~\ref{participant}).
Our recruitment strategy involved reaching out through company-internal channels and using snowball sampling~\cite{parker2019snowball} to identify suitable interviewees. 

We conducted the interviews  virtually, recorded them with consent, and transcribed them using an in-house tool. We followed the usual best practices for interviews~\cite{seidman2006interviewing,paul} to establish rapport and encourage open dialogue.

We analyzed the interview data with standard qualitative methods~\cite{hsieh2005three,lazar2017research}, involving open coding and memoing to develop and iteratively refine a codebook for challenges and solutions. We established inter-coder agreement on one interview (percentage agreement between two raters coding independently = 80\%).

\begin{table}[t]
\label{participant}
\caption{Interview Participants}
\begin{tabular}{@{}lll@{}}
\toprule
\textbf{Products} & \textbf{Participants} & \textbf{Roles} \\ \midrule
G1 & P01, P08, P18 & Manager, Engineer, Data Scientist \\
G2 & P03, P05, P06 & Engineers \\
G3 & P02, P9, P10, P17 & Manager, Data Scientist, Engineer \\
G4 & P04 & Engineer \\
G6 & P11 & Engineer \\
G7 & P12, P14 & Managers \\
G8 & P13, P24 & Researcher, Data Scientist \\
G9 & P15, P16 & Data Scientist, Manager \\
G10 & P19 & Manager \\
G11 & P21, P22, P23 & Engineers \\
G12 & P25, P26 & Manager, Data Scientist\\\bottomrule
\end{tabular}
\end{table}

\subsection{Survey}
Drawing from our interview analysis results, we identified the key challenges and emerging solutions and designed our survey to quantify their prevalence within the company, using various rating scales. This included questions about agreement to statements about challenges, ratings of difficulty of select activities (from `extremely hard` to `extremely easy`) and ratings of whether participants tried and used techniques (from `tried and would recommend' to `would like to try' to `tried but did not find useful'). Similar to the focus in our interviews, we again focused the survey on the quality assurance aspects of developing LLM-based products, allowing us to explore this topic in more depth while keeping the survey to a reasonable length. In addition, we also incorporated a separate section in our survey for non-LLM components to compare and contrast the responses — this is beyond the scope of this paper.

Since it is difficulty to identify who exactly works on LLM features at Microsoft, we oversampled and included anyone who committed to a repository containing LLM code as a potential survey participant. Despite expecting a lower response rate~\cite{punter2003conducting}, it was for us crucial to reach as many LLM practitioners as possible. To ensure accuracy, we incorporated qualifying questions into the LLM section of the survey. In total, we emailed 12,878 practitioners, received 977 automated out-of-office replies, and 332 responses (response rate $<$ 3\%), among which, 182 responses were directed to the LLM component of the survey.

In this paper, we report primarily quantitative results from ratings regarding challenges and emerging solutions.
 
\subsection{Threats to Validity and Credibility}
Our research has the kind of limitations typical for this style of research. Both of our interviews and surveys have a risk of response bias, where the respondents may provide inaccurate responses due to misunderstanding, social desirability, and other factors. While the survey affords some generalizability to the population of developers at Microsoft, given that participation in our survey and interviews was voluntary, those who chose to participate might be inherently different from those who did not, potentially skewing the results. Results might be influenced by other practices at Microsoft, hence readers should be careful when generalizing results to other organizations. Also despite following standard practices for coding and careful design of our survey and interview protocol, we cannot entirely exclude biases introduced by us researchers.

\section{Results}
We provided an overview of quality-assurance-related challenges identified in the literature and confirmed in our interviews and survey in Table~\ref{Table:challengesEval}. In the remainder of this paper, we focus on emerging solutions reported by the interviewed and surveyed practitioners, again focusing on quality assurance (we report other emerging solutions related to requirements, development and integration, and prompt development in the appendix).

\textbf{Challenges, disruptions, and emerging solutions:} Emerging solutions do not always match perfectly the identified challenges and many solutions implicitly or explicitly address multiple challenges. To organize the emerging solutions, we organize them by themes we call \textit{disruptions}. 
Specifically, we refer to \textit{challenge} as inherent difficulty or obstacles associated with and caused by LLMs and use \textit{disruptions} to describe how these challenges disrupt traditional software development practices and cause day-to-day disturbance for practitioners in their established workflows and practices (especially for practitioners new to LLMs) as they attempt to address these challenges within their known established practices and tools. The experienced disruption then drives the exploration and adoption of new solutions to overcome them. In essence, the sequence is: One or more challenges lead to disruptions in development practices, which in turn trigger the adoption of a new solution. For example, the challenge \textit{lack of specifications} (\challenge{C1}, Table~\ref{Table:challengesEval}) leads to perceived \textit{insufficiency of objective metrics}, which results in the formulation and adoption of the emerging practice of \textit{combining multiple qualitative and quantitative metrics} (Emerging Solution 2).

For each emerging solution, we include quantitative evidence from our survey (marked with \faBarChart). Based on the practitioners who rated that they (a) \textit{`tried and would recommend'} the solution, (b) \textit{`tried and found it somewhat useful,'} and (c)\textit{`tried but did not find it useful,'} we compute the percentages of respondents who \textit{adopted} as $a+b+c$, who was \textit{satisfied} as $a+b$, and who \textit{did not find the solution useful} as $c$.

\newcounter{disruptions}
\newcommand\disruption[2]{\refstepcounter{disruptions}\vspace{2pt}\subsection*{\textbf{Disruption \Alph{disruptions}: #1 (\challenge{#2}).}}}

\newcounter{solutions}
\newcommand\solution[2]{\refstepcounter{solutions}\vspace{2pt}\noindent\faGear\,\textbf{Emerging solution \arabic{solutions}: #1 \surveynumber{#2}.}}
\newcommand\solutions[4]{\refstepcounter{solutions}\vspace{2pt}\noindent\faGear\,\textbf{Emerging solutions \arabic{solutions} \& \refstepcounter{solutions}\arabic{solutions}: #1 \surveynumber{#2}. #3 \surveynumber{#4}.}}
\newcommand\solutionnodata[1]{\refstepcounter{solutions}\vspace{2pt}\noindent\faGear\,\textbf{Emerging solution \arabic{solutions}: #1 (no adoption statistics available).}}
\newcommand\solutioncustom[2]{\refstepcounter{solutions}\vspace{2pt}\noindent\faGear\,\textbf{Emerging solution \arabic{solutions}: #1 (#2).}}
\newcommand\solutionscustom[4]{\refstepcounter{solutions}\vspace{2pt}\noindent\faGear\,\textbf{Emerging solutions \arabic{solutions} \& \refstepcounter{solutions}\arabic{solutions}: #1 (#2). #3 (#4).}}

\disruption{Evaluation metrics change substantially}{C1, C2, C3, C4, C6}

Whereas metrics for test suite quality for traditional code (e.g., coverage) and metrics to evaluate qualities (e.g., response time, error rate) are fairly standardized and often automated, product teams at Microsoft often need to create custom metrics when evaluating LLM features and face substantial manual effort. For example, subjective qualities such as fluency, saliency, consistency, and creativity are considered crucial for generating text based suggestions in some product use cases (P1, P3, P5).
There is usually no clear oracle (not even the labels of traditional ML model testing), but many outputs may be equally correct or acceptable for a given input.
Many more subjective measures are open to interpretation and teams routinely rely heavily on manual human judgement, making the process time-consuming and laborious (\surveynumber{76.6} survey respondents mentioned spending manual effort). Also, even with human evaluators, ensuring objectivity and reliability can be difficult, and may introduce inconsistencies and biases~\cite{aroyo2015truth}.

Overall, developers often need to creatively explore new metrics rather than relying on established ones -- for example P4 mentioned, \textit{"I just created this. It's called the [-] metric, which looks at like 10 dimensions. You can score across each dimension and then a final score.''} Interviewees find metric creation challenging, as P25 pointed out, \textit{“it's really a social science problem more than a science problem.”} The process of defining custom metrics lacks a systematic foundation, and teams frequently encounter difficulty in defining the right metric for their specific use case, as P7 rightfully mentioned \textit{``it's just frustrating to come up with some scoring criteria.''}

\solutioncustom{Defining custom metrics through iterative collaboration and expert consultations}{\allsurveynumbers{33.1}{21.9}{25.8}{11.3}{6}{2}}
Product teams have started to use initial brainstorming sessions aimed at figuring out the various dimensions of quality related to their specific LLM use cases (P1, P4, P15, P16, P17, P22, P25, P26). Instead of approaching this as a typical engineering task, they treat it more as a research phase. Many teams engage with domain experts (e.g., \textit{linguists}, P1) and researchers (P15, P16, P17) to explore specific use cases and identify metrics for their desired response qualities. P16 also mentioned involving LLMs as a judge to evaluate the appropriateness of such metrics, and human judgement to validate it further: \textit{``Like asking the LLM as a judge metric to evaluate how good those types of [metrics] are. [...] We have a data scientist who works on this. There's quite a bit of iteration, and then also we worked with [the] PM and the feature crew to explain what we've done and then ask for feedback as well to iterate on.''} Teams also mentioned iterating over it multiple times to confirm they have the right metrics, such as \textit{``You have to run through this couple of times to make sure you have the right set of metrics''} (P1).

\solutioncustom{Combining qualitative and quantitative metrics to evaluate the multifaceted outputs effectively}{\allsurveynumbers{33.8}{19.9}{27.2}{13.9}{4.6}{0.7}} \label{sol:diversemetrics}
Most interviewed teams (P2, P4, P8, P9, P15, P16, P17, P18, P22, P26) have started combining subjective metrics (e.g., fluency) with `objective,' more mechanical, more easily quantifiable metrics (e.g., evaluating the syntax of a generated formula) to make evaluation more comprehensive. For example, with reference to evaluating LLM-generated conclusion slides,  engineering manager mentioned \textit{``We do both structured objective metrics and subjective metrics about how good is this slide,''} where quantitative criteria like bullet point count and sentence length are assessed alongside qualitative elements such as content-groundedness. This hybrid approach is echoed by other teams: \textit{“We kind of have a merge of two things. One is still this correctness [...] You could find different ways to sum the two columns, but there's generally a functionally correct answer. [...] There's on the other side [...] We don't have a spec [...] So in these cases it's more subjective to even measure the quality on the output''}.

\solutionscustom{Evaluating subjective metrics using LLM validators}{\allsurveynumbers{26}{21.3}{24}{20.7}{4.7}{3.3}}{Establishing clear rubrics and scoring mechanisms}{\allsurveynumbers{17.6}{15.5}{37.2}{20.9}{8.1}{0.7}} 
The usage of LLMs as validators for evaluating LLM responses has gained attention in the literature~\cite{Murugadoss:EvaluatingTheEvaluator:2024,zheng2023judging,shankar2024validates,kamoi2024evaluating} and has become a common practice for many teams at Microsoft (P1, P2, P3, P4, P6, P9, P10, P11, P15, P16, P17). These teams use rubrics to measure certain qualitative factors such as fluency (smoothness and proficiency of language), salience (contextual appropriateness), and consistency (uniformity and lack of contradictions) that signal the expected response quality of the model. Microsoft has developed frameworks \surveynumber{15.4} to facilitate this technique, garnering positive feedback from many, P1:\textit{``Today you can actually use large language models to evaluate the outcome of large language models, which is fantastic.''} Of the twelve product teams we engaged with, nine incorporate this approach in at least one feature team, highlighting its effectiveness in reducing manual efforts.

Despite these benefits, the approach comes with its potential pitfalls. One key reported issue among our participants appears to be the perceived unreliability of the validator LLMs. Some participants describe the validator LLMs as \textit{``flaky,''} suggesting that there are inconsistencies in the judgement delivered by these models for similar responses. This inconsistency makes it challenging to rely solely on these LLM validators for response evaluation, thereby necessitating the involvement of human validators. However, teams still found it beneficial to \textit{use LLMs for an initial validation phase, followed by a secondary review by human validators} to verify the initial pass/fail test results.

One big concern of this approach research has found is that using LLM-as-a-judge to directly evaluate a model's output can result in a model assigning a high score if it was likely to produce that output and not necessarily based on criteria~\cite{Murugadoss:EvaluatingTheEvaluator:2024}. As a result, an excessive dependency on these LLM validators could potentially lead to oversights of such issues by human validators, thereby instilling a misleading sense of validation accuracy. Our observations have also identified an unwarranted over-reliance on this subjective method across various teams. Notably, we found teams using LLM validators even in instances where adopting more objective measures could be better suited, such as for measuring the syntactical correctness of generated code. This finding underscores the need for thoughtfulness and careful consideration in selecting validation techniques (see \textit{emerging solution~\ref{sol:diversemetrics}}). 

\solutioncustom{Build a validator allowing a range of acceptable outputs instead of conducting single-valued unit tests for objective metrics}{\allsurveynumbers{29.1}{13.5}{26.4}{23.6}{6.8}{0.7}} 
To improve the robustness of their objective evaluation, many Microsoft teams adopted test designs that accept a range of multiple acceptable answers or entirely switch to test general criteria instead of expecting one acceptable value (P2, P9 P12) -- which mirrors ideas from property-based testing~\cite{claessen2000quickcheck,Ribeiro2023-rw}. They systematically implement this by using techniques such as regular expressions, search patterns, similarity algorithms, or simply by matching with a pre-approved set of cases. P12 mentioned their strategy of using a similarity algorithm: \textit{``We know what's a good answer; we can calculate the similarity of the generated answer to the good answer [using a comparison algorithm], and if the similarity is close, although there could be some variation, but if the similarity to the known good answer is high, then we feel the quality is high.''} Similarly, P2 mentioned using a formula evaluation pipeline for formula verification: \textit{``So whenever there's a formula that shows up, we make sure that the formula goes through an evaluation pipeline before it comes back.''} The notable benefit of such systematic evaluation techniques is that they can be largely automated, which substantially reduces the need for labor-intensive manual work.


\disruption{Common assumptions about test processes and environments break}{C5, C6, C9} 
Model and prompt evaluations use different infrastructure than traditional unit testing and rely often on large amounts of data that needs to be handled separately. And while developers are used to continuous integration for code tests, they do not necessarily set up similar infrastructure to (automatically) re-run offline evaluations whenever the model or prompt pipeline changes. 

Also, given all the differences and new challenges when it comes to LLMs (including the ones previously discussed), practitioners often do not have experience or templates of how to approach model and prompt testing, resulting in many ad-hoc processes. These ad-hoc processes tend to be inefficient and can result in practitioners dealing with repetitive, redundant tasks, causing frustration and suboptimal outcomes.

\solutioncustom{Automating offline evaluation to run periodically on a schedule}{\allsurveynumbers{19.5}{9.4}{37.6}{22.1}{10.7}{0.7}} 
A few teams (P4, P6, P22) we spoke to value the practice of periodically scheduled, automated offline evaluations. For example, P22 mentioned, \textit{``Most of the offline eval is heavily automated. I mean basically you wanna have a scheduled run every day just to keep track of the general metrics.''} P6 also voiced her interest in adopting such a routine, given she saw other teams benefiting from it: \textit{``They have [tests] just running on a schedule [..], like every hour or something. Evaluating sort of their offline response quality and they have a dashboard for that too, and I've asked some folks how can we do that because I think it's also valuable to have the offline evaluation happening periodically.''}

\solutioncustom{Establishing internal team standards for evaluation processes and pipelines}{\allsurveynumbers{32.2}{15.8}{29.6}{17.8}{3.9}{0.7}}
To overcome this, several teams try to establish internal \textit{standards} for the evaluation process or pipeline (P2, P4, P9, P15, P16, P22). As P15 expressed: \textit{``So as we were like developing these, we started facing challenges of like, OK, we can do this in an ad hoc way and we'll have to do this many times. So we need to start building things where we can do it at scale. [...] [Do not want to] just put data scientists on evaluating the same prompt on different models every time because that is kind of redundant. So that's where I think the thought for like standardization and like building standard pipelines and tools that came in.''} Beyond individual teams, there are efforts within Microsoft to provide default processes supported by standardized tools to benefit all product teams, as P11 stated: \textit{``So we're using fairly standardized tools that Microsoft other teams have provided us. [..] So what we need to do is hook up those services to our back end, which is fairly standard.''}

\disruption{Engineers require new skills to handle LLM evaluations}{C4}
Traditional software engineering education did not teach the skills necessary for evaluating LLMs (which did not exist even a few years ago). Such evaluations often require adopting a research-focused mentality where developers need to explore hypotheses, define custom metrics and measurements, and iterate over variations -- these are not commonly taught and practices in traditional software engineering. We observed that this shift can be challenging for engineers. For example, P9 explained how engineers lack the understanding that LLM evaluations are not similar to unit tests where all the test cases need to pass: \textit{``People who are trained doing research usually have the background of understanding that looking at the analysis resources is important. It's not a unit test, right? We often run into engineers thinking about these as unit tests. Say for example, this is a benchmark, right? It is OK that there is 63 failures. It just shows me that limits of the system [...] And engineers tend to think about it as ohh [...] I need to make the tests of unit test quality which is 100\% pass rate.''}

\solutioncustom{Involving data scientists in authoring tests, as they understand the limitations of the model better}{\allsurveynumbers{20}{12.1}{27.9}{27.9}{10.7}{1.4}} At Microsoft, teams (P7, P9, P10, P14, P15, P18, P26) often recruit explicit help from data scientists to author test cases, as P14 mentioned, \textit{``We have probably like 1 dedicated person from the data science team who's helping us kind of complete some of that testing.''} Interviewees find that involving data scientists in authoring prompts and test cases  improves the testing process. With their understanding of the model's capabilities and limitations, data scientists can more rigorously assess the data involved to generate the test cases, as P9 mentioned, \textit{``That's actually the place where I think the domain of data science is important. [...] really carefully thinking through the data.''}. They are often well-versed in spotting trends, anomalies, and potential pitfalls that may be overlooked by others. Interviewees believe that data scientists are better positioned to preempt potential issues and create test cases that cover a wider range of scenarios, leading to a more comprehensive LLM evaluation. Conversely, the absence of a data scientist in the team can prove to be a roadblock, as P5 explained: \textit{``We don't know what's going on and we don't know how to remedy this and we don't have a data scientist on the crew to under to basically explain what was going on.''} 

\disruption{Even with extensive evaluations, LLM solutions remain unreliable and developers have difficulty establishing trust}{C4, C5} 

Even with substantial evaluations and more systematic and automated processes, teams often still doubt the quality and reliability of their features, as P10 explained: \textit{``It's easy to just generate something, right? It's easy to have a solution and get up and running with it, [...]  It's a lot harder to really know, like where the quality is at [...] lot harder to kind of have confidence.''} The practitioners lack of confidence in LLM responses are not unfounded, as they have recounted numerous incidents that validate their concerns, as P5 mentioned: \textit{``There have been [...] multiple times where we've been caught off guard [...] like any non deterministic system, there would always be edge cases where [the LLM] just blurbs out it's entire command, which is a huge risk. [..] It's been on the news.''}
As a general theme, we see much more emphasis on testing and monitoring in production and on providing guardrails beyond the model.

\solutioncustom{Employing canary release strategy to enhance confidence in the LLM outputs}{standardized practice, no adoption statistics available}
All interviewed teams use canary releases. A canary release strategy is the idea expose a new feature or update initially to only few users and monitor the behavior of the system, before rolling it out to more users, or rolling it back if problems are discovered~\cite{releasenegineeringbook}.
Microsoft employs several stages of audience groups, called `rings,' through which a new release passes before becoming available to the public. As P3 explained: \textit{``We'll do [RING-2]. We'll keep it in internally for a while [...] before we actually go beyond. So there's kind of like a breaking in, burning in, stabilization period.''} This helps catch any unforeseen bugs or issues that might not have been apparent during development or initial testing stages. For instance, P10 discovered from the earliest release ring that the LLM was not performing up to par: \textit{``We had a [RING-1]. We had like an early release of the [LLM] experience, and it was not good. [...] It was pretty buggy.''} Beyond detecting bug, canary release also allowed interviewees to gather early user feedback on the new feature or update, that are used to make improvements or adjustments before a full deployment, P4: \textit{``So then once we get into Microsoft rings in production, we can look through [customer] feedback [...] and then figure out which were the right comments we wanna address.”} Beyond internal rings, Microsoft also has customers under non-disclosure agreements (NDAs) who participate in early access programs, use the product before others, and provide valuable feedback, as explained by P14: `\textit{`then we have our EAP, our early access program. And these are like a set of customers that we work very closely with. Those customers are very open about, this is the scenario we tried and it didn't work, and then we can use that feedback.”}

\solutioncustom{Running A/B testing for tracking changes in different versions (e.g., prompt updates, and model migration)}{\allsurveynumbers{25.4}{11.3}{32.4}{22.5}{8.5}{0}} 
P22 appreciated the utility of online A/B testing frameworks, stating, \textit{``I think the online AB test framework, I think it's pretty nice and it's pretty detailed.''} Several teams (P8, P12, P16, P18, P21, P22, P24, P26) also mentioned their consistent use of A/B testing for different types of changes such as change of underlying model, and updates in the prompt engineering pipeline, such as P18: \textit{``during the migration we run AB testing like 50-50\% and try to compare the two models directly.''}  P16 also mentioned running audits on the A/B testing to make sure its serving its purpose: \textit{``On our the AB testing and metrics we're doing this audit to understand if the telemetry is implemented correctly, because that's one area that we want to make sure things are done well in too.''}

\solutionscustom{Establishing extensive guardrails}{\allsurveynumbers{21}{20.3}{25.4}{21}{9.4}{2.9}}{Monitor systems to trigger alerts automatically if something goes wrong}{\allsurveynumbers{21.1}{13.1}{31.4}{26.3}{5.8}{2.2}} 
Some teams (P1, P2, P3, P4, P8, P10, P11, P16, P17, P22) at Microsoft have implemented robust guardrails, including API format checks, response structure validations, specific pattern regular expressions, and other rule-based checks. This has improved practitioners' confidence in the overall system, P8: \textit{``I dont see a risk like the content would be able to mess up with the engineering system. Reason being we have put a lot of good guardrails to make sure that the reliabilities are good enough [...] all like prompt injection, check block list and like safety harm, we call that content safety, and couple of mores like jailbreak classifiers.''} 

However, there can occasionally be a false sense of security from these guardrails. For instance, P11 mentioned the three-level guardrail system, falsely considering the trained model itself as the first guardrail, \textit{``LLM is usually a already trained or fine tuned to to to not answer to an appropriate questions So it's already what you call censored''}, instructions to the model as the second guardrail, \textit{``the second thing we have a control over is the metaprompt. So you we can say please answer politely. Please don't answer any legal or medical questions and so on''}, and finally the real guardrail, \textit{``The third one is [tool] policy filters. So the the input and the output are run through a filter. They will look to see if there's any bad words and so on and so forth.''}

Despite these misconceptions, there's an impression that guardrails have improved over time. Practitioners (P3, P7) also highlighted the development of systems to automatically alert them if issues arise, assuring improved protection and monitoring. As P3 pointed out, \textit{``It's gotten better, guardrails over the years. There are systems now that know the prompt you sent in and watch for the prompts coming back out in the result and can can automatically trigger or recycle or can tell you that it couldn't get an answer or can do things like that. So we have more protections to these kinds of things coming out.}"

Microsoft also maintains a framework with universal guard lists, which set uniform standards across all its products while also reducing the individual teams' workload in establishing guardrails. However, one concern we found among a few practitioners is how these guardrails can sometimes be too sensitive and flag innocuous interactions with customers, which can lead to customer frustration, as P15 mentioned \textit{``we are not able to understand the sensitivity of the guardrails that we are building [...] I remember in January or December of this year we got our feedback [...] somebody said that yeah, [feature] was overblocking. [...] So Power user wrote on Reddit that like I mean actually they wrote a blog like that.''} 

\disruption{Existing approaches to telemetry and monitoring need to be revised}{C3, C9} Many of the interview participants mentioned that the current telemetry methods, originally developed for non-ML software, are not strong indicators when it comes to assessing the quality of LLM-generated responses. As P8 elaborated: \textit{``I don't think we have any kind of signal which indicate that we are doing an awesome job on generating content.''}. Many teams depend on the classic telemetry metrics, such as direct customer feedback, thumbs up, and thumbs down, which typically have a very low response rate.

A particularly frequently mentioned problem is no-eyes debugging, that is, to gain insights into system behavior at runtime without accessing user data, which should not be revealed to developers due to compliance and privacy regulation. This problem exists in debugging traditional software systems too~\cite{zeller2009programs}, but is reported as even more severe with the open-ended nature of interacting with LLMs where users can input any text and the models can produce any response. Developers struggle to find better signals about when problems occur private in LLM outputs (e.g., as opposed to crash logs).
For example, P13 explained, \textit{``All of the commercial data like [LLM responses] are eyes off. So we cannot actually see it. So how you go and like understand what people are doing and what's working and how to make it all better if I can't actually even see it [...] that's like a huge, huge challenge and I think just in general, you know, Microsoft is super careful with customer data, [...] so you just have to do the best you can while keeping you know sort of customer data privacy.''}

\solutioncustom{Developing new and multiple types of telemetry metrics that may better suit LLM solutions}{\allsurveynumbers{43.2}{14.2}{24.3}{12.2}{5.4}{0.7}}
Practitioners employ several types of telemetry to assess the performance of their LLM features. For example, some practitioners (P7, P17, P20, P24, P25, P26) track \textit{apology rates} from LLMs, an indicator of the models' confidence in their responses. There is another set of metrics that almost all Microsoft practitioners use predominantly, called \textit{seen-tried-kept}, that measures whether customers noticed the feature, tried it, and accepted or used the generated response. Moreover, Microsoft have a custom telemetry metric called ASHA (Aggregated Session HAppiness) to estimate customer satisfaction, referred as the \textit{happiness index}. This measures the success rate of each user sessions, such as, if a user employs the LLM feature ten times without failure, they have a successful ASHA session. P17 elaborated the metrics they use: \textit{``We have standard KPIs, [...] we use Asha, which sort of this is for aggregated session happiness scores. So we use that pretty heavily and we rely a lot on the customer feedback thumbs up and thumbs down. [...] beyond that, there's like seen tried and kept, of course [..] also say an engagement measurement of do they continue to engage with the [feature]. And we also have a measure of apologies like how often we're apologizing in a response, right. So that's another indication that we're probably not giving high confidence answers.''} Apart from all these, they still have other traditional metrics such as monthly active users (P15).

Research teams in Microsoft are also actively engaged in establishing a robust set of quality metrics for the product teams, largely based on the analysis of system logs, as P13 mentioned: \textit{``the sort of mission in terms of coming up with a set of measures based on people's naturalistic interactions with [LLMs] in order to give guidance to the product team. [...] They track of course a bunch of key metrics, one of which is percent of conversations that were successful. And so how we created that measure with them was what indicator that metric [...] we call inferred SAT, inferred user satisfaction. Every single [LLM] interaction gets a score''}

\solutioncustom{Use the LLM-as-a-validator strategy to gain granular insight in production behavior without revealing private data}{\allsurveynumbers{11.3}{12.8}{37.6}{28.4}{9.2}{0.7}}
Instead of solely relying on established but weaker signals like \textit{seen} or \textit{kept rates}, a few teams are trying to identify and design stronger signals for their use cases. For instance, P17 mentioned an approach where validator LLMs are used to review the eyes-off user conversations and report back with an assessment of the quality:\textit{``So there's one path we have which compliant eyes off chat analysis [..] where an LLM is able to look over [user-LLM conversation] in an anonymized fashion and basically report back if it's done well, if the conversations are good or not [...] So that's again relying on LLM evaluators and eyes off chat analysis. It's one way we try to get signal.''} P6 also mentioned using a similar approach to track whether the quality of responses is improving or deteriorating over time: \textit{``We have online [LLM-as-judge] evaluation, which is sort of like evaluating the actual conversations [...] eyes off, so it doesn't actually save the evaluation, but it's sort of evaluates it at the time it's happening. And then it saves that, and we have this sort of daily chart that shows us how is the quality of these real responses happening or changing over time.''} 

P15's team went a further step up to validate whether the metric delivers accurate signals by showing the LLM's evaluation to the user and asking for their agreement or divergence: \textit{``So we had built in an evaluation within the product itself. The first few versions of the product where the user was able to look at the rating that was given by the LLM and then select if they agree or disagree with that rating and how much rating they would give it. And then submit logs. [...] those logs are shared back. You don't read the customer content, but you know a certain [response]’s model score and user score, and if there is a difference, it means like the prompt engineering is not [working well].''} Additionally, P15's team is running a similarity analysis algorithm to track the variation between the text generated and the text ultimately used by the user: \textit{``What was the output from the LLM? And then what was the final [text] that was [used]? So the plan is to do text distance analysis on the output of the LLM to the final [text] that was [used], because then you can actually understand how much of a variation was coming in terms of like the final [text] content.”} 

\disruption{Lack of focus on system-wide evaluation of LLM-based products}{C5} 
The new challenges with evaluating LLMs tend to draw attention to model and prompt evaluation approaches (forms of unit testing) and away from more holistic evaluations of the entire system (integration, system, and acceptance testing).

\solutioncustom{Setting up an end-to-end test automation infrastructure}{\allsurveynumbers{22.7}{14.9}{33.8}{20.1}{7.8}{0.6}} 
Recognizing the need for comprehensive system testing to assess the full functionality and efficiency of LLMs in the system, Microsoft's infrastructure teams have developed an end-to-end testing framework. Being a consumer of the framework, P17 explained the need for it: \textit{``We're evaluating the LLM; we likely wanna evaluate the full system, the end to end system, because the client [program] is a part of our system that does non trivial computation as a part of our conversation with our LLM. There are plenty of great tools to evaluate a single LLM call or maybe like a couple of tightly interconnected LLM calls [...] but as soon as we execute an [client program] and that gets fed into prompt, the tooling support doesn't exist. So [infrastructure team] having to create our own.''} Although this platform is relatively new, many Microsoft product teams (\surveynumber{39.4}) have already adopted it.

\solutioncustom{Conducting comprehensive tests beyond unit tests, including tests for reliability and availability}{\allsurveynumbers{39}{21.3}{19.1}{13.5}{6.4}{0.7}}
Beyond evaluating LLMs, Microsoft teams also prioritize regular software testing. This includes conducting unit and regression tests, among others, which are integral in identifying and rectifying bugs throughout the software system. For instance, P2 noted the importance of unit tests in preventing issues: \textit{``the app teams have a very extensive like unit test pipeline. So like it's just unit testing at the end of the day. As long as our tests aren't failing, then we're not breaking anything else in the product.''} Similarly, P11 referred to regression testing: \textit{``if we change the [filters], if we change our meta prompt, you know anything that's substantial or the large language model itself, if you switch from 3.5 to four, then you will have to rerun the test.”}, P7 mentioned UX testing: \textit{``We also have some UX tests or end to end tests on validate that the the answers are coming as expected.''}, and P3 discussed reliability testing: \textit{``We can measure [reliability]. We wanna get 99\%. Then you go until you hit 99\%. It's a very clear goal.''}

\disruption{Responsible AI, being a relatively new concept for engineers, has a steep learning curve and might come across as bureaucratic}{C9}
\label{RAI}
Practitioners often consider Responsible AI as a secondary priority~\cite{rakova2021responsible,nahar2023meta}, particularly when they are already burdened with a substantial workload to develop, integrate, and evaluate such products with models. However, improper prioritization and a lack of standardized Responsible AI (RAI) practices can lead to legal, ethical, and reputational risks for organizations. Microsoft places a high priority on responsible AI practices~\cite{MSFTRAI}, incorporating ethical considerations, transparency, and human-centric values into their development processes to ensure the responsible deployment of AI technologies, including LLMs. 

\solutionnodata{Standardizing Responsible AI (RAI) evaluation and practices -- bias and fairness evaluation \surveynumber{39.7}, safety evaluation \surveynumber{55.9}, robustness evaluation \surveynumber{33.5}, transparency and explainability \surveynumber{27.9}, privacy compliance \surveynumber{55.8}, security and vulnerability evaluation \surveynumber{59.8}, other \surveynumber{11.7}} As P7 highlighted, \textit{``It feels like right now it's we are doing heavy [RAI] testing around that area just because we are new, we don't know what's gonna happen. We just want to play safe.''} This statement emphasizes that Microsoft is placing a significant emphasis on RAI assessment. The company has assembled dedicated teams and specialists who focus on Responsible AI practices, particularly for large language models these days. They employ different types of RAI checks, P21: \textit{``We basically what we call responsible AI filtering where we do some sort of separate checks to make sure that we're not outputting sort of racist or sexist or other comment things that shouldn't, that totally inappropriate to be saying.''}, and at different phases, P8: \textit{``We do post filter and prefilter on the input content and the output content for responsible AI''}. These experts also establish the necessary metrics, measurements, and tool support required for evaluating Responsible AI, which all teams adhere to, As noted by P3, \textit{``So there's a RAI responsible AI stack that catches if it's anything harmful.''} Now that they have established and standardized these practices, they are focusing on automation and aiming to increase the frequency of RAI activities, P4: \textit{``We're trying to do that more frequently. So we're looking at how do we run RAI just weekly and there's some new tools that many other teams are building to make that happen automatically.''} The process and the automation are complicated as P17 described, and so the experts are helping the feature teams go through and use them effectively, \textit{``There are many redundant tools all running and trying to catch inappropriate content at different layers, so helping teams integrate that into their system. The other side would be like understanding the requirements and consulting I guess with feature crews, [...] making sure we understand the prompts, the metaprompts, the mitigations that are in place''}

\solution{Applying RAI red teaming strategies}{48}
Red teaming, in general, is a technique where experts mimic cyber-attacks to identify vulnerabilities in a company's security system. Responsible AI (RAI) red teaming is focused on detecting and rectify any flaws related to Responsible AI within AI systems, to ensure responsible AI practices are enforced. P1 elaborates on how Microsoft has extensively enhanced its RAI red teaming process, transforming it from random tests to a more sophisticated system, \textit{``red teaming used to be very random, right? Just to try ourself and give it a few prompt and see what triggers something offensive. Now it is more sophisticated [...] sometimes we also invite people from the bigger org to participate [...] so we can get more diverse inputs and observations.''} P17 builds on this by detailing the layered structure of this technique in the company: \textit{``There's a few different layers that does red teaming, so at the [upper team] side, there are some security red teaming that's done and they've also dipped their toes in some like, RAI harms red teaming as well. At that, the high level and the lower down on the [team], there is an [product] specific we call it kind of a purple team: but it is the red team that also have internal knowledge of the system.”} P22 also shared about an existing effort that combines offline evaluation with red teaming within one tool and a single codebase to simplify the process: \textit{``So actually one of the things we've done recently is we moved all of our red teaming from [local repo] into our offline evaluation tooling. So it's all the same code base, making it easier for when you have offline eval you get red team, because both are generally needed to get to any kind of release.''}

\solutioncustom{Following a robust RAI audit process}{mandatory for adoption} At Microsoft, it is mandatory for all teams to gain approval from an organizational entity known as the Deployment Safety Board (DSB) prior to releasing any feature related to LLMs. The DSB's primary role is to reaffirm that all teams have adhered to standard Responsible AI checks and practices. As P12 elaborated: \textit{``Yeah, we run DSB, Privacy, we do all sort of review and validation before we release. So that's also why the release cycle has to be impacted. It just a lot of testing, a lot of evaluation and signoff has to happen before we release each update or each feature. But yeah, we don't want customers to experience harmful content.''} In addition to initial approval, teams are also required to re-submit to the DSB for each significant change such as model migration. This constitutes a comprehensive process as P21 explained, \textit{``Every time we change the model we're using [...] have to rerun all that offline evaluation and then do red teaming and get DSB approval and then we can do A/B and to see how that looks.”}

\section{Discussion and Conclusion}

This study provides insights into how practitioners are addressing the disruptions they face when  integrating LLMs into software products. By identifying 19 emerging solutions, we have highlighted strategies that are already being widely adopted to manage the complexities of LLMs, particularly in the areas of quality assurance. These practices—such as defining custom evaluation metrics, combining qualitative and quantitative measures, and automating offline evaluations—offer practical solutions for handling LLM-specific disruptions.

Our findings have important practical implications. For development teams, adopting practices such as automating offline evaluations, involving data scientists in test creation, and using LLMs as validators can improve the development and evaluation of LLM-based features. These solutions not only help teams manage the unpredictable and subjective nature of LLMs but also provide a roadmap for integrating these new processes into existing workflows.

While the widespread adoption of many of these solutions by teams at Microsoft, and the frequent explicit endorsement in the survey, implicitly suggests effectiveness, we have not conducted a rigorous evaluation of how well they work in different contexts or compared them to potential alternatives. A logical next step in this line of research (for us or others) is to carry out such a study to assess the effectiveness of these solutions in various settings. As LLMs become more prevalent, development teams will continue to face these, and likely even more, disruptions.  Our aim is to provide a set of validated best practices that teams can confidently follow to ensure that these features will behave as intended, without going off track or causing unintended outcomes.
We encourage other researchers and practitioners to share their experiences and insights as they adopt solutions, so that the community of software professionals collectively can learn how to safely and responsibly develop LLM-enabled software products.

\section*{Acknowledgment}
We thank all our interview and survey participants and those who helped us find and recruit them. We also thank Chenyang Yang and Jenny Liang for their feedback and suggestions, especially about prompt engineering and testing research.

\bibliographystyle{ieeetr}
\bibliography{references}
\flushend

\end{document}